\documentclass[12pt]{iopart}

\usepackage{iopams}
\usepackage[utf8]{inputenc}
\usepackage[english]{babel}
\usepackage{color}
\usepackage{ulem}
\usepackage{lineno}

\newcommand{\be}{\begin{equation}}
\newcommand{\ee}{\end{equation}}
\newcommand{\beq}{\begin{eqnarray}}
\newcommand{\eeq}{\end{eqnarray}}
\newcommand{\bk}{\bigskip\noindent}
\newcommand{\sk}{\smallskip\noindent}
\newcommand{\n}{\nonumber}

\def\keywords#1{\vspace{10pt}
     \begin{indented}
     \item[]\rm Keywords: #1\par
     \end{indented}}
     
\def\AMS#1{\vspace{10pt}
     \begin{indented}
     \item[]\rm AMS classification scheme numbers: #1\par
     \end{indented}}

\begin{document}

\title[Hamilton-Jacobi approach for Regge-Teitelboim cosmology]{Hamilton-Jacobi
approach for Regge-Teitelboim cosmology}

\author{Alejandro Aguilar-Salas$^1$, Alberto Molgado$^{2,3}$ and Efra\'\i n 
Rojas$^{4}$}

\address{$^1$ Facultad de Matem\'aticas, Universidad Veracruzana, 
91000, Cto. Gonz\'alo Aguirre Beltr\'an s/n, Xalapa, Veracruz, 
M\'exico}

\address{$^2$ Facultad de Ciencias, Universidad Aut\'onoma de San Luis 
Potos\'{\i} Campus Pedregal, Av. Parque Chapultepec 1610, Col. 
Privadas del Pedregal, San Luis Potos\'{\i}, SLP, 78217, Mexico}

\address{$^3$ Dual CP Institute of High Energy Physics, Colima, Col, 28045, Mexico} 

\address{$^4$ Facultad de F\'\i sica, Universidad Veracruzana, 91000, 
Cto. Gonz\'alo Aguirre Beltr\'an s/n, Xalapa, Veracruz, M\'exico}

%\address{$^5$ Departamento de F\'\i sica, Escuela Superior de F\'\i sica 
%y Matem\'aticas, Instituto Polit\'ecnico Nacional, Unidad Adolfo L\'opez Mateos, 
%Ciudad de M\'exico, M\'exico}

\ead{alej.aguilar.salas@gmail.com, alberto.molgado@uaslp.mx, efrojas@uv.mx}

\vspace{10pt}

% \begin{indented}
% \item[]November, 2020
% \end{indented}

\begin{abstract}
The Hamilton-Jacobi formalism for a geodetic brane-like 
universe described by the Regge-Teitelboim model is 
developed. We focus on the description of the complete 
set of Hamiltonians that ensure the integrability of 
the  model in addition to obtaining the Hamilton 
principal function $S$. In order to do this, we avoid 
the second-order in derivative nature of the model by 
appropriately defining a set of auxiliary variables that 
yields a first-order Lagrangian. Being a linear in 
acceleration theory, this scheme unavoidably needs an 
adequate redefinition of the so-called Generalized 
Poisson Bracket in order to achieve the right evolution 
in the reduced phase space. Further this Hamilton-Jacobi 
framework also enables us to explore the quantum behaviour 
under a semi-classical approximation of the model. A 
comparison with the Ostrogradski-Hamilton method for 
constrained systems is also provided in detail.  
\end{abstract}



\AMS{70H20, 70H45, 70G75, 81T30, 83F05}

%%%%%%%%%%%%%%%%%%%%%%%%%%%%%%%%%%%%%%%%%%%%%%%%%%%%%%%%%%%%%%%%%%%%%%%%%%%%%%
\section{Introduction}
\label{sec:1}
%%%%%%%%%%%%%%%%%%%%%%%%%%%%%%%%%%%%%%%%%%%%%%%%%%%%%%%%%%%%%%%%%%%%%%%%%%%%%

The Regge-Teitelboim (RT) model, also named geodetic brane 
gravity, deals with our 4-dimensional universe as an extended 
object geode\-sically floating in a higher-dimensional flat 
Minkowski spacetime~\cite{RT1975}. The underlying motivation 
of this model was to develop General Relativity (GR) following 
the principles that ordinarily are used to determine the behaviour 
of either the worldline of a particle or the worldvolume of an 
extended object by considering the embedding functions instead 
of the metric as the fundamental geometric objects. 
The theory exhibits a built-in Einstein limit which sets this 
model as an attractive theoretical tool that deserves further 
development. Both the RT model and GR have in common that the 
action that describe them depend linearly on second-order 
derivatives of the fields variables that, respectively, characterize 
them. Within the mini-superspace brane-like cosmology framework,  
such dependence is more evident and manageable. 
This last issue has been worked at greater length at both 
the classical and the quantum levels in many contributions~\cite{davidson1998,davidson2003,
ostro2009,paston2010,paston2012,biswajit2013,pavsic2001}.

From the viewpoint of the Dirac-Bergmann approach for constrained 
systems~\cite{dirac1964,henneaux1992,rothe2010}, the RT model
is described by a singular Lagrangian system that has been 
analyzed by using different strategies, all of them interesting 
in their own. However, there is a variational alternative to study 
the aforementioned singular nature of the theory based on the 
\textit{equivalent Lagrangians method} introduced by 
Carath\'eodory~\cite{caratheodory1967} for regular systems, and 
further developed for the treatment of singular systems  by G\"{u}ler \cite{guller1992a,guller1992b}, and shortly afterwards extended 
to higher-order derivative theories by Pimentel and collaborators \cite{pimentel1996,pimentel1998,pimentel2005,pimentel2008}.
Even though this approach has as a starting point the Lagrangian scheme, 
it brings into play a suitable shortcut to the Hamilton-Jacobi (HJ) 
framework without going through the usual approximation of the 
canonical transformations of the Hamiltonian formalism even in 
the singular case. Indeed, within this geometrical HJ framework 
the constraints emerge as a set of partial differential equations 
(PDE) that must obey certain conditions that ensure their 
integration. Contrary to the situation in the Dirac-Bergmann 
approach, in this formalism it is not mandatory to classify the 
constraints as first- and second-class constraints. In fact,
in this HJ approach a singular physical system is viewed as a 
many variables system which leads to replace the usual Hamilton 
equations by a set of total differential equations in those 
variables. On mathematical grounds, for singular systems it is 
not possible to solve such equations unless we meet certain 
geometrical conditions known as \textit{integrability conditions} 
which could lead to more constraints. Certainly, one of the 
central roles of this approach is to complete the analysis on 
the integrability of singular systems through the analysis of 
the conditions under which the canonical field equations are 
integrable or not. The canonical equations naturally appear as 
total differentials expanded in terms of the parameters of the 
theory, and whenever these form a fully integrable set, 
their simultaneous solutions will determine the generating 
function $S(\tau, z^A)$, the Hamilton principal function, uniquely 
by imposing some initial conditions.

The purpose of this paper is to offer a novel geometrical 
alternative to analyze the integrability of the geodetic
brane gravity within the minisuperspace cosmological scenario. 
This is performed in terms of the HJ framework. As we will see, 
in order to construct the HJ structures for the RT model, it is 
convenient to introduce auxiliary variables in order to reduce 
the complexity of the analysis by considering a first-order in 
derivatives Lagrangian. As a consequence of this extension in 
the configuration variables, the number of integrability 
conditions to be studied increases. Nevertheless, the purpose 
behind the reduction of the order is twofold. First, to use a 
HJ approach developed for first-order actions, derived from 
higher-order derivative theories and thus apply the standard 
HJ method and, second, to detect the local gauge symmetries of 
the theory through a purely geometrical approach in order to 
achieve the correct gauge transformations. Furthermore, unlike 
the existing distinct Hamiltonian approaches, we claim that we 
are able to obtain the Hamilton principal $S$-function, which in 
turn will allow us to explore its quantum semi-classical 
approximation. Indeed, the knowledge of this $S$-function has a 
direct implication when one tries to obtain some information on 
quantum singular systems by investigating the semi-classical WKB 
approximation. In this case, the constraints are promoted to 
conditions that the wave function must satisfy at the semi-classical 
limit, in addition to satisfy the Schr\"{o}dinger equation.

The layout of the paper is as follows. In section 
2 we provide an overview of the HJ formalism, adapted to 
first-order actions. In section 3 we implement this formalism
for the RT model, adapted to a FRW geometry, by introducing 
auxiliary variables. We elucidate in detail the 
conditions under which the integrability conditions emerge by 
making contact with the Ostrogradski-Hamiltonian approach. 
We also address the local gauge symmetries for this model. 
In section 4, we obtain the Hamilton principal function and 
briefly discuss the semi-classical approximation of the 
quantum approach for the RT cosmological model. Conclusions 
are presented in section 5. 

%%%%%%%%%%%%%%%%%%%%%%%%%%%%%%%%%%%%%%%%%%%%%%%%%%%%%%%%%%%%%%%%%%%%%%%%%%%%%%%%%
\section{Hamilton-Jacobi framework for first-order actions}
\label{sec:2}
%%%%%%%%%%%%%%%%%%%%%%%%%%%%%%%%%%%%%%%%%%%%%%%%%%%%%%%%%%%%%%%%%%%%%%%%%%%%%%%%

Consider that the dynamical behaviour of a physical system 
is described by the action
\begin{equation} 
\mathcal{S}[z^A] = \int d\tau \, L (z^A, \dot{z}^A, \tau),
\qquad \qquad
A,B = 1,2,,3,\ldots, N;
\label{eq1}
\end{equation}
where $z^A$ are the coordinates of the associated $N$-dimensional 
configuration manifold $\mathcal{C}_N$, and $\dot{z}^A$ denote 
their associated velocities. Here an overdot stands for the 
derivative with respect to the time parameter $\tau$. Since 
our interest lies in cosmological models with a linear dependence 
on accelerations~\cite{affine2016}, we first proceed to construct 
a HJ framework for first-order actions i.e., theories with a linear
dependence in the velocities, and then we will implement this 
for the RT cosmological model. The explicit form of the Lagrangian 
function to be considered is
\be 
L (z^A,\dot{z}^A, \tau) = K_A (z^B)\,\dot{z}^A - V (z^B).
\label{eqL}
\ee
Summation over repeated indices is henceforth assumed. The 
variational principle selects the optimal trajectory $z^A = z^A 
(\tau)$ parametrized by $\tau$. Here, $K_A (z)$ and $V(z)$ are 
assumed to be smooth functions defined on $\mathcal{C}_N$. The 
Euler-Lagrange equations of motion (eom) are 
\be 
\label{eom0}
\left( \frac{\partial K_B}{\partial z^A} - \frac{\partial 
K_A}{\partial z^B}\right)\dot{z}^B - \frac{\partial V}{\partial 
z^A} = 0 .
\ee

Following \textit{Carath\'eodory's equivalent Lagrangians 
approach}~\cite{caratheodory1967,guller1992a,guller1992b}, in 
order to have an extreme configuration of the action~(\ref{eq1}) 
the necessary and sufficient conditions are associated to 
the existence of a family of surfaces defined by a generating 
function or Hamilton's principal function, $S(z^A,\tau)$, 
such that it satisfies
\beq 
\frac{\partial S}{\partial z^A} = \frac{\partial L}{\partial 
\dot{z}^A} =: p_A,
\label{eq2a}
\\
\frac{\partial S}{\partial \tau} + \frac{\partial S}{\partial 
z^A} \dot{z}^A - L = 0,
\label{eq2b}
\eeq
where $p_A$ denotes the conjugate momenta to the coordinates 
$z^A$. Note that in our particular case the conjugate momenta 
is given by $p_A = K_A (z^B)$.

\bk
The HJ framework in which we are interested arises from~(\ref{eq2a}) 
and~(\ref{eq2b}) considered as partial differential equations 
(PDE) for the generating function $S$. Indeed, for non-singular 
systems it is relatively straightforward to obtain the function 
$S$ by expressing the velocities $\dot{z}^A$ in terms of the 
coordinates $z^A$ and partial derivatives of $S$ what is provided 
by appropriately inverting Eq.~(\ref{eq2a}). However, for singular 
physical systems this may not be as direct, as we will see below. 
Even worse, for affine in velocity (or acceleration) 
theories~\cite{affine2016} this turns out rather intricate as the 
Hessian matrix of the system vanishes identically,
\be 
H_{AB} = \frac{\partial^2 L}{\partial \dot{z}^A \partial 
\dot{z}^B} = 0.
\label{eq3}
\ee
Clearly, for the Lagrangian we are considering, the rank of 
the Hessian matrix is zero which causes the phase space to be 
non-isomorphic to the tangent bundle of the configuration 
manifold, $T^*\mathcal{C}_N$. This means that the manifold 
$\mathcal{C}_N$ is fully spanned by the $R = N - 0 = N$ variables 
where the $z^A$ play the role of parameters from the HJ viewpoint~\cite{guller1992a,guller1992b}. Indeed, these parameters are 
related to the null space of the Hessian $H_{AB}$. Of course, 
we can not invert the velocities $\dot{z}^A$ in 
favor of the coordinates and partial derivatives of the function 
$S$, that is, $\dot{z}^A \neq f^A \left( \tau, z^B, \partial 
S/\partial z^B \right)$. 

In what follows, it is convenient to introduce the notation 
\be
t^A := z^A 
\qquad \qquad \mbox{and} \qquad \qquad
H_A :=  - \frac{\partial L}{\partial \dot{t}^A} = - K_A (z^B),
\label{def1}
\ee
which, by relation~(\ref{eq2a}), follows 
\be 
\frac{\partial S}{\partial t^A} + H_A \left( \tau, t^B, 
\frac{\partial S}{\partial t^B}\right) = 0.
\label{eq5}
\ee
Notice that $H_A$ does not depend on $\dot{t}^A 
= \dot{z}^A$. Similarly, taking into account~(\ref{def1}) and by 
introducing the Hamilton function
\be 
\label{H0}
H_0 := \frac{\partial S}{\partial t^A} \dot{t}^A - L(\tau,t^A, 
\dot{t}^A),
\ee
one finds that the expression~(\ref{eq2b}) becomes
\be 
\label{eq7}
\frac{\partial S}{\partial \tau} + H_0 \left( \tau, t^B, 
\frac{\partial S}{\partial t^B}\right) = 0,
\ee
which is most commonly known as the Hamilton-Jacobi equation.

\sk
Now, we are able to collect~(\ref{eq5}) and~(\ref{eq7}) into a 
single equation expressing a unified set of PDE  for the 
generating function $S$. To perform this, we simply introduce 
the notation $t^0 := \tau$, and thus we find
\be 
\label{eq8}
\frac{\partial S}{\partial t^I} + H_I \left( t^J, 
\frac{\partial S}{\partial t^A}\right) = 0, \qquad \qquad
I, J = 0,1,2,\ldots, N.
\ee
In the following, these $N+1$ relations will be referred to as the 
\textit{Hamilton-Jacobi partial differential equations} (HJPDE). 
Bearing in mind~(\ref{eq2a}), it is very useful to express
the HJPDE in the Hamiltonian fashion
\be 
H_I ' (t^J, p_J) := p_I + H_I (t^J, p_J) = 0,
\label{canonC}
\ee
where we have considered $p_0 := \frac{\partial S}{\partial \tau}$.
These relationships have thus acquired the well-known form 
of canonical Dirac constraints. In other words, to get a 
clearer picture, this HJ approach replaces the analysis of the 
$N$ canonical constraints, $H_A ' = 0$, with the analysis of 
the $(N+1)$ HJPDE given by relations~(\ref{eq8}).

Within this framework the equations of motion are 
written as total differential equations. These are often known 
as the \textit{characteristic equations} associated to the 
Hamiltonian set~(\ref{canonC}). In our particular case these 
characteristic equations are given by 
\beq 
dz^I = \frac{\partial H_J '}{\partial p_I} dt^J, 
\label{dz1}
\\
dp_I = -\frac{\partial H_J '}{\partial z^I} dt^J.
\label{dp1}
\eeq
At this initial stage we must notice that all coordinates $z^A$ 
have a status of independent evolution parameters, in a similar 
fashion as the time parameter $\tau$. On mathematical grounds, 
within this HJ formalism it is said that $t^I = z^I$ are the 
\textit{independent variables} or \textit{parameters} of the 
theory. On physical grounds, this may be confusing but we may 
consider that these parameters encode the local symmetries and 
gauge transformations, as we will see below. We have thus enlarged 
the configuration space to be $\mathcal{C}_{N+1}$. Moreover, 
another important equation is provided by the 
function $S$. Indeed, we have that
\be 
dS = \frac{\partial S}{\partial \tau} d\tau + \frac{\partial 
S}{\partial z^A}dz^A = - H_I \,dt^I,
\label{dS0}
\ee
where~(\ref{def1}) and~(\ref{canonC}) have been considered. 

\sk
For two arbitrary functions defined on the cotangent bundle
of the configuration manifold, $F, G \in \Gamma_{N+1} := 
T^* \mathcal{C}_{N+1}$, that is, functions in the extended 
phase space spanned by the variables $z^I = (t^0, t^A)$ and 
their conjugate momenta $p_A = (p_0, p_A)$, we introduce the 
\textit{extended Poisson bracket} (PB)
\be 
\{ F, G \} = \frac{\partial F}{\partial z^I}\frac{\partial 
G}{\partial p_I} -  \frac{\partial F}{\partial p_I}\frac{\partial 
G}{\partial z^I}, \qquad \qquad
I, J = 0,1,2,\ldots, N.
\label{EPB}
\ee
We may therefore express evolution in $\Gamma_{N+1}$ as follows
\be
dF = \{ F, H_I '\}\,dt^I,
\label{dF}
\ee
where the role that the $t^I$ play as parameters of the Hamiltonian 
flows generated by the constraints $H_I '$ is more evident. It is 
worth mentioning that the characteristic equations~(\ref{dz1}) 
and~(\ref{dp1}) also may be obtained from~(\ref{dF}). In this HJ 
framework, the dynamical evolution provided by~(\ref{dF}) is 
referred to as the \textit{fundamental differential},~\cite{guller1992b}.

%%%%%%%%%%%%%%%%%%%%%%%%%%%%%%%%%%%%%%%%%%%%%%%%%%%%%%%%%%%%%
\subsection{Integrability conditions}
\label{subsec:2a}
%%%%%%%%%%%%%%%%%%%%%%%%%%%%%%%%%%%%%%%%%%%%%%%%%%%%%%%%%%%%

With the intention of integrating the HJPDE~(\ref{canonC}), it is 
convenient to rely in the method of characteristics~\cite{caratheodory1967}.
On physical grounds, it is not clear whether or not  all coordinates 
are relevant parameters of the theory, so it is crucial to find 
a subspace among the parameters where the system becomes integrable. 
Regarding this, the matrix occurring in~(\ref{eom0}) 
\be 
M_{AB} := \{ H_A ', H_B ' \} = \frac{\partial K_B}{\partial 
z^A} - \frac{\partial K_A}{\partial z^B},
\label{MAB1}
\ee
enters the game in order to unravel under what conditions 
first-order actions will have integrability. Geometrically, 
this matrix may be interpreted as the ``curl'' of the vector $K_A$.  

\sk
The complete solution of~(\ref{canonC}) is given by a family
of surfaces orthogonal to the characteristic curves. The
fulfillment of the \textit{Frobenius integrability 
conditions}~\cite{caratheodory1967,pimentel2014},
\be 
\{ H' _I, H' _J \} = C^K _{IJ}\,H'_K,
\ee
ensures the existence of such a family where $C^K_{IJ}$ are the 
structure coefficients of the theory. This means that the 
Hamiltonians must close as a Lie algebra. Hence, it must be 
imposed that both $dH_0 '$ and $dH_A '$ are vanishing identically
\be 
d H_I ' = 0.
\label{integrability}
\ee

\sk
In view of this, we shall discuss the possible 
scenarios~\cite{pimentel2005,pimentel2008}
\begin{itemize}
\item
First, if both $\{ H_0 ', H_A '\} = 0$ and $M_{AB} = 0$ 
identically, we will have $dH_I ' = 0$ so that $dt^I$ are 
independent. Accordingly, the equations of motion are all 
integrable.
\item
Second, $M_{AB} \neq 0$ and $\det (M_{AB}) \neq 0$, that is,
the regular case. The realization of $dH_I ' = 0$ leads to
consider that $dt^0$ and $dt^A$ are dependent. In such case
it is often enough to consider that $t^0 = \tau$ is the 
independent parameter of the theory and the evolution of 
$F \in \Gamma_{N+1}$ is provided by
\be 
\label{dF1}
dF = \{ F, H_0 ' \}^*\,dt^0,
\ee
where
\be 
\{ F, G \}^* := \{ F, G\} - \{ F, H_A ' \} (M^{-1})^{AB}
\{ H_B ', G\}.
\label{DB1}
\ee
Here, $(M^{-1})^{AB}$ denotes the inverse matrix of $M_{AB}$
such that $M_{BC}(M^{-1})^{CA} = \delta^A _B$ or $(M^{-1})^{AC}
M_{CB} = \delta^A _B$.
\item 
Third, $M_{AB} \neq 0$ but $\det (M_{AB}) = 0$, that is,
the singular case. The rank of $M_{AB}$ being, say $P = N - R$,
implies the existence of $R$ null eingenvectors $\lambda^A _{(\alpha)}$, 
or \textit{zero-modes}, of $M_{AB}$ such that $M_{AB}
\lambda^B _{(\alpha)} = 0$ with $\alpha = 1,2,\ldots, R$. 
This fact causes a split of the manifold $\mathcal{C}_N$ 
in two submanifolds: $\mathcal{C}_R$ spanned by $R$ 
coordinates, $t^\alpha = z^\alpha$ related to the kernel 
of $M_{AB}$ and $\mathcal{C}_P$ spanned by $P$ coordinates, 
namely $z^{A'}$, with $A' = R+1,R+2, \ldots, N$, associated 
to the regular part of $M_{AB}$. 
Clearly, the fact that $P\neq 0$ leads to the existence of 
a $P \times P$ submatrix of $M_{AB}$, say $M_{A' B'}$, 
such that $\det (M_{A' B'}) \neq 0$ indicates the existence 
of an inverse matrix $(M^{-1})^{A' B'}$ satisfying 
$M_{B'C'}(M^{-1})^{C'A'} = \delta^{A'} _{B'}$ or $(M^{-1})^{A'C'} 
M_{C'B'} = \delta^{A'} _{B'}$. Therefore, the condition 
$dH_I ' = 0$ causes the presence of $t^{\alpha'}$ independent 
parameters such that the evolution is given by
\be 
\label{dF2}
dF = \{ F, H_{\alpha'} ' \}^*\,dt^{\alpha'}, 
\qquad\qquad
\alpha' = 0,1,2,\ldots, R;
\ee
where 
\be 
\{ F, G \}^* := \{ F,G\} - \{ F, H_{A'} ' \} (M^{-1})^{A'B'}
\{ H_{B'} ', G\}.
\label{DB2}
\ee
In this HJ spirit, the remaining variables $z^{A'}$ are
referred to as \textit{dependent variables}.
In passing, in this particular
case, the condition $dH_A ' = 0$ yields
\be 
C_{(\alpha)} := \frac{\partial H_0}{\partial z^A}\, 
\lambda^A _{(\alpha)} = 0,
\label{condition1}
\ee 
where the explicit value of $\{ H_A ', H_0'\}$ has been used.
This orthogonality condition among the zero-modes and 
$\mathcal{H}_A := \partial H_0/ \partial z^A$, 
induces a very convenient strategy to identify 
a totally equivalent set of constraints to those emerging
in the integrability procedure behind this HJ formalism 
adapted for first-order 
actions. Likewise, these \textit{Lagrangian constraints} 
can be obtained straightforwardly from the eom~(\ref{eom0}).
\end{itemize}
The new bracket structure, introduced either in (\ref{DB1}) 
or in (\ref{DB2}), is referred to as the \textit{generalized 
Poisson bracket} (GPB) which has all the known properties 
of the standard Poisson bracket. This redefine the dynamics 
by eliminating some of the coordinates with exception of the 
$t^{\alpha'}$. As a matter of fact, this structure is closely 
related to the Dirac bracket arising in the Dirac-Bergmann 
Hamiltonian approach for constrained systems~\cite{dirac1964,
henneaux1992,rothe2010}. On the other hand, it may happen that 
the constraints $H_I ' = 0$ do not satisfy the condition $dH_I ' 
= 0$ identically when~(\ref{dF1}) or~(\ref{dF2}) are 
considered as fundamental differentials. In such case, this 
condition leads us to obtain equations of the form $f(z^A, p_A) 
= 0$ which should also be considered as constraints of the 
system. These in turn, must obey the established integrability 
condition, $df = 0$, which can lead to more constraints.  
Once all the constraints have been found, it is mandatory to 
incorporate them within the HJ framework where some of them 
must be considered as generators of the dynamics. It should be 
remarked that this incorporation must be accompanied by the 
introduction of more parameters to the theory, these related 
to the new constraints that generate dynamics, derived from the 
imposition of integrability. Therefore, the space of parameters 
has been increased where, every arbitrary parameter is in 
relation to the generators of the dynamics~\cite{pimentel2008,
pimentel2014}. Accordingly, in this new scenario, the fundamental 
differential becomes modified once the complete set of parameters 
has been identified and incorporated to the theory
\be 
dF = \{ F, H'_{\overline{\alpha}} \}^*\,dt^{\overline{\alpha}}.
\label{dF3}
\ee
Here, $t^{\overline{\alpha}}$ denotes the complete set of 
independent parameters of the theory, where the index 
$\overline{\alpha}$ covers the entire set of these parameters. 
As a result, the fundamental differential~(\ref{dF3}) must be 
used to obtain the right evolution in the reduced phase space 
through the GPB. 
In a like manner, consistency of conditions~(\ref{condition1}) 
requires to analyse the relation $d C_{(\alpha)} = \{ C_{(\alpha)}, 
H_{\alpha'}'\}^* \, dt^{\alpha'} =: \mathcal{C}_{(\alpha,\alpha')}\, 
dt^{\alpha'} = 0$. This could bring further conditions of the type 
$\mathcal{C}_{(\alpha,\alpha')} (z,p) =0$ that should be also 
treated as constraints of the theory on an equal footing to the 
former set of constraints. One must continue with this iterative 
algorithm until no further independent constraints emerge.

\bk
Hereinafter, unless otherwise stated, we confine ourselves 
to the third case above. On the technical side, the matrix 
$(M^{-1})^{A'B'}$ defines a reduced symplectic structure on 
the phase space $T^* \mathcal{C}_{N+1}$, where $M_{AB}$ is singular. 
This structure provides the appropriate dynamics of the 
system as we can observe from~(\ref{dF2}).
Regarding this, when considering $F=z^{A'}$ in (\ref{dF2}), 
it is straightforward to show that 
\be 
dz^{A'} = (M^{-1})^{A' B'} \left( \frac{\partial K_{B'}}{\partial 
z^{\alpha'}} - \frac{\partial K_{\alpha'}}{\partial z^{B'}} \right)
\,dt^{\alpha'}, 
\qquad \alpha' = 0, 1,2,\ldots, R;
\label{charac2}
\ee
where one must consider the definition $K_0 := - H_0$ with $H_0$ 
given by~(\ref{H0}). In arriving to these characteristic equations 
we have used the explicit value of $\{ H'_{A'} , H'_{\alpha'}\}$. 
We would like to stress some points. First, from the latter 
expression we readily infer that the solutions to~(\ref{charac2}) 
will be of the form $z^{A'} = z^{A'} (t^{\alpha'})$, which represent
congruences of $(R+1)$-parameter curves where the $t^{\alpha'}$ 
play the role of parameters. Second, by inserting~(\ref{charac2}) 
into~(\ref{dS0}) one finds
\be 
\label{dS1}
dS = -\left[  H_{\alpha'} + H_{A'} (M^{-1})^{A' B'} \left( 
\frac{\partial K_{B'}}{\partial z^{\alpha'}} - \frac{\partial 
K_{\alpha'}}{\partial z^{B'}} \right)\right] \,dt^{\alpha'},
\ee
where $H_A '$ is introduced by~(\ref{def1}). Third, the 
characteristic equations~(\ref{charac2}) will be considerable simplified 
for the case of the RT gravity as a consequence of  the affine 
in acceleration property of the model, as we will see shortly.

%%%%%%%%%%%%%%%%%%%%%%%%%%%%%%%%%%%%%%%%%%%%%%%%%%%%%
\subsection{Generator of gauge symmetries}
\label{subsec:2b}
%%%%%%%%%%%%%%%%%%%%%%%%%%%%%%%%%%%%%%%%%%%%%%%%%%%

In reference~\cite{pimentel2014} it has been argued that once 
the complete set of involutive Hamiltonians $H'_{\overline{\alpha}} 
= 0$ of the theory has been found, i.e., $\{ H'_{\overline{\alpha}}, 
H'_{\overline{\beta}} \}^* = C^{\overline{\gamma}}_{\overline{\alpha} 
\overline{\beta}}\,H'_{\overline{\gamma}}$, these Hamiltonians 
must be considered as generators of infinitesimal canonical 
transformations in $T^* \mathcal{C}_{P+1}$ as follows
\be 
\delta z^A = \{ z^A, H'_{\overline{\alpha}} \}^*\,\delta 
t^{\overline{\alpha}}.
\label{flows1}
\ee
These are referred to as the \textit{characteristic flows} of the 
system. Here, $\delta t^{\overline{\alpha}} := \bar{t}^{\,\,
\overline{\alpha}} - t^{\overline{\alpha}} = \delta t^{\overline{\alpha}} 
(z^A)$. In particular, when one set $\delta t^0 = 0$, 
expression~(\ref{flows1}) defines a special class of transformations
\be 
\delta z^A = \{ z^A, H'_{\dot{\alpha}} \}^*\,\delta t^{\dot{\alpha}},
\label{flows2}
\ee
which, by imposing that they remain in the reduced phase space, 
$T^* \mathcal{C}_P$, form the so-called infinitesimal contact 
transformations in the spirit of the constrained Hamiltonian 
framework by Dirac,~\cite{dirac1964}. In this sense, they do not 
alter the physical states of the system. In~(\ref{flows2}), 
$t^{\dot{\alpha}}$ denotes the set of all independent parameters 
where $t^0$ is excluded. Clearly, the transformations~(\ref{flows2}) 
are generated by 
\be 
G := H' _{\dot{\alpha}} \,\delta t^{\dot{\alpha}},
\label{G}
\ee
so that~(\ref{flows2}) is equivalent to
\be 
\delta_G z^A = \{ z^A, G \}^*.
\label{flows3}
\ee
Thus, $\delta_G z^A$ is the specialization of~(\ref{flows1}) to  
$T^* \mathcal{C}_P$ where $G$ is the generating function of 
the infinitesimal canonical transformation. In the spirit of 
the theory of gauge fields, transformations~(\ref{flows3}) set 
the gauge transformations of the theory.

%%%%%%%%%%%%%%%%%%%%%%%%%%%%%%%%%%%%%%%%%%%%%%%%%%%%%%%
\section{RT cosmological brane theory}
\label{sec:3}
%%%%%%%%%%%%%%%%%%%%%%%%%%%%%%%%%%%%%%%%%%%%%%%

The original RT gravity, including a cosmological constant term
$\Lambda$, is described by the action~\cite{RT1975}
\be 
\mathcal{S} [X^\mu] = \frac{\alpha}{2} \int_m d^{3+1}x\,
\sqrt{-g}\,\mathcal{R} - \int_m d^{3+1}x\,\sqrt{-g}\,\Lambda,
\label{action}
\ee
where $X^\mu$, the embedding functions, are the field 
variables describing the $4$-dim trajectory, $m$, spanned by 
a 3-dim extended object in its evolution in a flat Minkowski
spacetime with metric $\eta_{\mu\nu}$, $\mu,\nu = 0,1,2,\ldots, 4$;
$\mathcal{R}$ stands for the Ricci scalar defined on $m$, 
$g= \det (g_{ab})$ with $g_{ab} = \eta_{\mu\nu}\, 
\partial_a X^\mu \partial_b X^\nu$ being the induced metric
with $a,b = 0,1,2,3$, and $\alpha$ is a constant with appropriate 
dimensions. 
%Here, we will consider additionally a cosmological constant 
%term $L_m \simeq \Lambda$ living on the brane.

\sk
Under the parametrization $x^\mu = X^\mu (\tau,\chi,\theta,\phi)
= (t(\tau),a(\tau),\chi,\theta,\phi)$ and by assuming that 
the background spacetime is expressed by $ds_5 = - dt^2 + da^2
+ a^2 d\Omega^2_3$ with $d\Omega^2_3$ being the unit three-sphere
metric, the induced metric gets the FRW geometry  
$ds_4 = - N^2 d\tau^2 + a^2 d\Omega^2_3$. Accordingly,
within the mini-superspace framework the action~(\ref{action}) 
reduces to $\mathcal{S} = 6\pi^2 \int d\tau \, L (t,a,\dot{a}, 
\dot{t},\ddot{a},\ddot{t},\tau)$
where the Lagrangian is explicitly given by~\cite{ostro2009}
\be 
L (t,a,\dot{a},\dot{t},\ddot{a},\ddot{t},\tau)= \frac{a \dot{t}}{N^3} 
\left( a \dot{t} \ddot{a} - a \dot{a} \ddot{t} + N^2 \dot{t} \right)
- N a^3 \bar{\Lambda}^2.
\label{lag0}
\ee
Here $\bar{\Lambda}^2 := \Lambda/3\alpha$ is a constant,
and $N := \sqrt{\dot{t}^2 - \dot{a}^2}$ represents the 
lapse function that commonly appears when we perform an 
ADM decomposition of the RT action~(\ref{action}),~\cite{ostro2009}. 
Further, the dot stands 
for derivative with respect to the parameter $\tau$.
Notice that this is a second-order derivative theory where now
$t$ and $a$ span the configuration space. The Hessian matrix 
with respect to the second-order derivatives associated to the 
Lagrangian~(\ref{lag0}) vanishes identically
\be 
H_{bc}:= \frac{\partial^2 L}{\partial \ddot{x}^b \partial 
\ddot{x}^c} = 0, \qquad \qquad b,c = 1,2.
\ee
This defines what is usually known as an affine in acceleration 
theory,~\cite{affine2016}. It is important to mention that for 
this Lagrangian one may identify a total derivative term by 
considering the decomposition
\be
L = L_1 (t, a,\dot{t},\dot{a},\tau) 
+ L_2 (t,a,\dot{t},\dot{a},\ddot{t},\ddot{a},\tau),
\label{lag01}
\ee
where
\beq 
L_1 &:=& - \frac{a \dot{a}^2}{N} + aN \left( 1 - a^2 H^2\right),
\label{lag1}
\\
L_2 &:=& \frac{d}{d\tau} \left( \frac{a^2 \dot{a}}{N} \right).
\label{lag2}
\eeq
It is quite evident that accelerations $\ddot{t}$ and $\ddot{a}$ 
enter in the Lagrangian~(\ref{lag0}) only through the total 
derivative term~(\ref{lag2}), and thus the equations of motion 
describing the system will be of second-order.

%%%%%%%%%%%%%%%%%%%%%%%%%%%%%%%%%%%%%%%%%%%%%%%%%
\subsection{Auxiliary variables}
\label{subsec:3a}
%%%%%%%%%%%%%%%%%%%%%%%%%%%%%%%%%%%%%%%%%%%%%%%%%

When treating the velocities and accelerations as 
independent variables the introduction of Lagrange 
multipliers is required to get a Lagrangian 
with only first-order derivatives. This allows us to 
implement the HJ framework developed above. 
The obvious choice is 
\beq 
x^b _s &= \{ x^b ; X^b \} = \{ t, a; \dot{t}, \dot{a} \},
\qquad \qquad \quad b = 1,2 \qquad s=0,1;
\label{xs}
\\
v^b &= \{ \dot{X}^b \} = \{ \ddot{t}, \ddot{a} \},
\label{vs}
\eeq
where the index $b$ ranges over the number of field 
variables while the index $s$ keeps track of the order of 
the derivatives of the field variables. Explicitly, 
\be 
\begin{array}{lll}
x^1 = t & \quad  X^1 = \dot{x}^1 =: T & \quad \quad\Longrightarrow
\qquad \dot{x}^1 - X^1 = 0,
\\
x^2 = a & \quad X^2 = \dot{x}^2 =: A & \quad \quad \Longrightarrow
\qquad \dot{x}^2 - X^2 = 0,
\\
v^1 = \dot{X}^1= \ddot{x}^1  =: \mathcal{T} & &\quad  \quad
\Longrightarrow \qquad \dot{X}^1 - v^1 = 0,
\\ 
v^2 = \dot{X}^2 = \ddot{x}^2  =: \mathcal{A} & & \quad \quad
\Longrightarrow \qquad \dot{X}^2 - v^2 = 0.
\end{array}
\label{id2}
\ee

\sk
The introduction of~(\ref{id2}) into~(\ref{lag0}), 
$L \rightarrow L^{(x_s,v)} = L^{(x_s,v)} (x^1,x^2,X^1,X^2,v^1,v^2)$, 
yields
\beq 
L^{(x_s,v)} &= - \frac{a^2 T}{\mathcal{N}^3}(A v^1 - T v^2) + 
\frac{aT^2}{\mathcal{N}} - \mathcal{N}a^3 \bar{\Lambda}^2.
\label{L0}
\eeq
Henceforth, for the sake of simplicity we shall use the short 
notation $\mathcal{N}:= \sqrt{T^2 - A^2}$ for the lapse function. 
In order to have a well defined Lagrangian, it is mandatory to 
introduce a set of Lagrange multipliers enforcing the 
constraints~(\ref{id2}), say
\be
\pi^s _b \equiv \{ \pi _b ; \Pi_b \}
= \{ \pi_1, \pi_2; \Pi_1, \Pi_2 \}.
\label{multipliers0}
\ee
Thus, the extended Lagrangian $L_E = L_E (x^a_s, v^a,\pi_a ^s)$ 
replacing the RT cosmological brane-like model~(\ref{lag0}) is  
given by
\be
\fl
L_E = L^{(x_s,v)} (x^a_s,v^a) + \pi_1 (\dot{x}^1 - X^1) 
+ \pi_2 (\dot{x}^2 - X^2) + \Pi_1 (\dot{X}^1 - v^1)
+ \Pi_2 (\dot{X}^2 - v^2).
\label{lagE0}
\ee
From this new viewpoint, we have an enlarged 10-dimensional 
configuration space, $\mathcal{C}_{10},$ spanned by the
variables $\{ x^a_s,v^a, \pi^s_a\}$. Specifically, 
the extended Lagrangian~(\ref{lagE0}) reads
\be 
L_E 
= \pi_1 \dot{x}^1 + \pi_2 \dot{x}^2 + \Pi_1 \dot{X}^1
+ \Pi_2 \dot{X}^2 - H^{(x_s,v,\pi^s)} (x_s^a,v^a,\pi^s_a),
\label{lagE1}
\ee
where
\be
H^{(x_s,v,\pi^s)} (x_s^a,v^a,\pi^s_a) = \pi_1 X^1 + \pi_2 X^2 
+ \Pi_1 v^1 + \Pi_2 v^2 - L^{(x_s,v)},
\label{ham0}
\ee
and the Lagrangian $L^{(x_s,v)}$ is defined by~(\ref{L0}). We infer 
that this function is the canonical Hamiltonian associated to this 
cosmological brane-like model, now described by the extended 
Lagrangian $L_E$ but, with the peculiarity that at this level 
it solely depends on the coordinates $z^A$ and not on their 
conjugate momenta.

\sk
As we already mentioned, the purpose of extending the 
configuration space is to bring the Lagrangian~(\ref{lag0}) 
into the form provided by~(\ref{eqL}) in order to implement 
a HJ analysis of it as a first-order model. To do this, 
we suitably choose the following ordered coordinates 
with the aim to make transparent the relation to the 
first-order Lagrangian in $\mathcal{C}_{10}$, 
\be
\hspace{-5ex} 
z^A = \{ v^a, x^a_s, \pi^s _a \} = \{ \mathcal{T}, 
\mathcal{A},t,a,T,A,\pi_t,\pi_a,\Pi_T,\Pi_A \}, 
\hspace{3ex} A = 1,2,\ldots,10.
\label{zA}
\ee
We are able to readily identify the values for $K_A$ 
from the Lagrangian~(\ref{lagE1})
\be 
K_A = ( K_\alpha , K_{A'} ) = 
\{ 0,0, \pi_1, \pi_2, \Pi_1, \Pi_2, 0,0,0,0 \},
\label{KA}
\ee
which allows us to explicitly split the values for the $A$ 
index: $A = (\alpha, A')$, where $\alpha = 1,2$ and $A' = 
3,4, \ldots, 10$. In the same spirit, $V$ turns to 
$H^{(x_s,v,\pi^s)}$ as given by~(\ref{ham0}). In passing, from~(\ref{eom0}),~(\ref{lagE1}),
(\ref{zA}) and~(\ref{KA}), the solely eom of this theory reads
\be
\mathcal{E}:= \frac{d}{d\tau} \left(\frac{A}{T} \right) 
+ \frac{\mathcal{N}^2}{a T} \frac{\Theta}{\Phi} = 0,
\label{eom1}
\ee
where 
\beq
\Theta &:= T^2 - 3 \mathcal{N}^2 a^2\bar{\Lambda}^2,
\label{Theta}
\\
\Phi &:= 3T^2 - \mathcal{N}^2 a^2\bar{\Lambda}^2.
\label{Phi}
\eeq

\sk
As dictated by~(\ref{canonC}), the HJPDE of our theory 
are explicitly given by
\be 
\label{hamiltonians}
H_I ' = \cases{
H_{\alpha'} ' = \cases{
H_0 ' = p_\tau + H^{(x_s,v,\pi^s)} = 0,
\\
H_{v^1} ' = p_\mathcal{T} = 0,
\\
H_{v^2} ' =  p_\mathcal{A} = 0,
}
\qquad \quad \alpha' = 0,1,2;
\\
H_{A'} ' = \cases{
H_t ' = p_t - \pi_1 = 0,
\\
H_a ' = p_a - \pi_2 = 0,
\\
H_T ' = p_T - \Pi_1 = 0,
\\
H_A ' = p_A - \Pi_2 = 0,
\\
H_{\pi_1} ' = p_{\pi_t} = 0,
\\
H_{\pi_2} ' = p_{\pi_a} = 0,
\\
H_{\Pi_1} ' = p_{\Pi_T} = 0,
\\
H_{\Pi_2} ' = p_{\Pi_A} = 0.
}
\qquad \qquad \quad A' = 3,4,\ldots, 10.
}
\ee
Note that $H_0 '$ is the only constraint depending
on the $v^a$ coordinates, that is, on the coordinates 
associated to the second-order derivatives.
Further, one may straightforwardly check that the only 
non-vanishing PB among the $H_A '$ are
\beq 
\nonumber
\{ H_3 ' , H_7 ' \} &= \{ H_t ',H_{\pi_1} '\} = -1 
\qquad \qquad 
\{ H_5 ' , H_9 ' \} \,\,\,= \{ H_T ', H_{\Pi_1} ' \} =-1,
\\
\nonumber
\{ H_4 ' , H_8 ' \} &= \{ H_a ' ,H_{\pi_2} '\} = -1 
\qquad \qquad 
\{ H_6 ' , H_{10} ' \} = \{ H_A ',H_{\Pi_2} '\} =-1,
\eeq
and thus we can construct the matrix~(\ref{MAB1}) 
\be 
( M_{AB} )= \left(
\matrix{
0_{2\times 2} & 0_{2\times 2} & 0_{2\times 2}
\cr
0_{2\times 2} & 0_{4\times 4} & - I_{4\times 4}
\cr
0_{2\times 2} & I_{4\times 4} & 0_{4\times 4}
}
\right) \qquad \qquad A,B = 1,2, \ldots, 10.
\label{matrix1}
\ee
The rank of $M_{AB}$ being $P= 8$ implies the existence 
of 2 null-eigenvectors, $\lambda^A_{(\alpha)}$,\, 
$\alpha = 1,2.,$ as well as an invertible 
submatrix and its associated inverse given by
\be 
\hspace{-5ex}
( M_{A'B'} )= \left(
\matrix{
0_{4\times 4} & - I_{4\times 4}
\cr
I_{4\times 4} & 0_{4\times 4}
}
\right) 
\quad \,\,\,\mbox{and}\,\,\, \quad 
(M^{-1})^{A'B'}= 
\left(
\matrix{
0_{4\times 4} & I_{4\times 4}
\cr
-I_{4\times 4} & 0_{4\times 4}
}
\right),
\label{matrix2}
\ee
respectively. Clearly, $\det \,(M_{A' B'}) \neq 0$. Notice 
that this is a symplectic matrix belonging to the symplectic 
group $Sp(8)$. 

\sk
In this HJ point of view the coordinates $z^A$ are separated in 
two sets
\be
\hspace{-4ex}
t^{\alpha} = \{ v^a \} = \{ \mathcal{T}, \mathcal{A} \}
\quad \mbox{and}\quad 
t^{A'} = \{ x^a_s, \pi_a ^s \} = \{ t,a,T,A,\pi_t ,\pi_a, 
\Pi_T, \Pi_A \}\,,
\label{zA2}
\ee
where $t^{A'}=z^{A'}$ are the true dynamical variables 
while $t^{\alpha'}$ are the parameters of the theory. 
Therefore, by expanding~(\ref{charac2}), and observing 
from~(\ref{KA}) that none of the $K_{A'}$ depend on the 
$t^{\alpha'}$, the characteristic equations transform into
\be 
dz^{A'} = (M^{-1})^{A'B'} \frac{\partial H^{(x_s,v,\pi^s)}}{\partial 
z^{B'}} dt^0,
\label{charact3}
\ee
while the remaining characteristic equation~(\ref{dS1}) reads
\be 
dS =  - \left[ H^{(x_s, v,\pi^s)} + H_{A'} (M^{-1})^{A'B'}  
\frac{\partial H^{(x_s, v,\pi^s)}}{\partial z^{B'}} \right]dt^0.
\label{dS3}
\ee
The characteristic equations~(\ref{charact3}) look like the 
standard Hamilton equations of motion, as expected, where 
$(M^{-1})^{A' B'}$ plays the role of a symplectic structure. 
Even though the characteristic equations~(\ref{charac2}) should 
depend, in principle, on the three parameters $t^{\alpha'}$, we 
realize that for our model the characteristic 
equation~(\ref{charact3}) dictates that, at this stage, 
$t^0$ is indeed the only relevant parameter.
Hence, the characteristic equations~(\ref{charac2}) have been 
replaced by the simplified set of equations~(\ref{charact3}). 
On physical terms, the only real parameter results to 
be $\tau$ while $\mathcal{T}$ and $\mathcal{A}$ are 
arbitrary, that is, they are pure gauge. This completely 
results as a consequence of the fact that these parameters 
belong to the second-order nature of our model which, however, 
may be effectively avoided by considering decomposition~(\ref{lag01}).

%%%%%%%%%%%%%%%%%%%%%%%%%%%%%%%%%%%%%%%%%
\subsection{Integrability analysis}
\label{subsec:3b}
%%%%%%%%%%%%%%%%%%%%%%%%%%%%%%%%%%%%%%%%%%%%

We proceed to analyse the integrability conditions on 
the Hamiltonians $H_I '$,~(\ref{hamiltonians}). 
The evolution of $H'_I$ dictated by~(\ref{dF2}) with the 
GPB~(\ref{DB2}), conveniently denoted at this stage by $\{ F, G\}^\bullet$ 
for reasons that will be clarified later, when using $(M_{A'B'})$ 
and its inverse given by~(\ref{matrix2}), leads to
the only non-vanishing terms
\beq 
dH_0 ' &= \left( \Pi_1 - \frac{\partial L^{(x_s,v)}}{\partial 
v^1} \right) dt^1 + \left( \Pi_2 - \frac{\partial 
L^{(x_s,v)}}{\partial v^2} \right) dt^2,
\n
\\
dH_{v^1} ' & = - \left( \Pi_1 - \frac{\partial 
L^{(x_s,v)}}{\partial v^1} \right) dt^0,
\n
\\
dH_{v^2} ' & = - \left( \Pi_2 - \frac{\partial 
L^{(x_s,v)}}{\partial v^2} \right) dt^0,
\n
\eeq 
By imposing the conditions 
$d H_I '  = \{ H'_I, H'_{\alpha'} \}^\bullet \,dt^{\alpha'} = 0$ 
we readily identify two new Hamiltonian constraints
\beq 
h_1 ' & := \Pi_1 - \frac{\partial L^{(x_s,v)}}{\partial v^1} 
= \Pi_1 + \frac{a^2 TA}{\mathcal{N}^3} = 0,
\label{H1}
\\
h'_2 & : = \Pi_2 - \frac{\partial L^{(x_s,v)}}{\partial v^2} 
= \Pi_2 - \frac{a^2 T^2}{\mathcal{N}^3} = 0,
\label{H2}
\eeq
where $L^{(x_s,v)}$, given by~(\ref{L0}), has been used. 
On the other hand, as discussed in section~\ref{sec:2}, 
the orthogonality condition~(\ref{condition1}) is totally 
equivalent to the integrability conditions~(\ref{H1}) 
and~(\ref{H2}). In this sense, we strategically opt to consider 
the orthogonality condition in order
to continue with the integrability analysis. To achieve 
this, we must compute the basis of the kernel of the 
matrix~(\ref{matrix1}) and then, construct suitable zero-modes.  
Firstly, we find that
\beq 
\label{uAs1}
u^A _{(1)} &= 
\left(
\matrix{
1 & 0 & 0 & 0 & 0 & 0 & 0 & 0 & 0 & 0
}
\right),
\\
u^A _{(2)} &= \left(
\matrix{
0 & 1 & 0 & 0 & 0 & 0 & 0 & 0 & 0 & 0
}
\right),
\label{uAs2}
\eeq
span a basis for the null space of the matrix $M_{AB}$. Accordingly, 
by considering
\be 
\lambda^A _{(1)} := X^1\,u^A_{(1)} + X^2\,u^A_{(2)}.
\label{lambda1}
\ee
The internal product between $\lambda^A_{(1)}$ and 
$\mathcal{H}_A:=\partial H_0/\partial z^A$, guided 
by~(\ref{condition1}), yields
\beq
\fl
\frac{\partial H^{(x_s,v,\pi^s)}}{\partial v^1} X^1
+ \frac{\partial H^{(x_s,v,\pi^s)}}{\partial v^2} X^2 
&= \Pi_1 X^1 + \Pi_2 X^2 -  \frac{\partial L^{(x_s,v)}}{\partial v^1} X^1
-  \frac{\partial L^{(x_s,v)}}{\partial v^2} X^2
= 0.
\label{id3a}
\eeq
Now, from~(\ref{L0}) we readily obtain the identity
$(\partial L^{(x_s,v)}/\partial v^1) X^1 + (\partial 
L^{(x_s,v)}/\partial v^2) X^2 = 0$. Thus, when substituting 
this into~(\ref{id3a}), one deduces directly
\be 
\label{C1a}
C_1 = \Pi_1 X^1 + \Pi_2 X^2 = 0.
\ee
In a like manner, by constructing the zero-mode
\be 
\lambda^A_{(2)} := X^2 \,u^A_{(1)} + X^1\,u^A_{(2)},
\label{lambda2}
\ee
the internal product between $\lambda^A_{(2)}$ and 
$\mathcal{H}_A$, following~(\ref{condition1}), reads
\be
\fl
\frac{\partial H^{(x_s,v,\pi^s)}}{\partial v^1} X^2
+ \frac{\partial H^{(x_s,v,\pi^s)}}{\partial v^2} X^1 
= \Pi_1 X^2 + \Pi_2 X^1 -  \frac{\partial L^{(x_s,v)}}{\partial 
v^1} X^2 -  \frac{\partial L^{(x_s,v)}}{\partial v^2} X^1
= 0.
\label{id4a}
\ee
As before, from~(\ref{L0}) it is fairly easy to verify
the identity $(\partial L^{(x_s,v)}/\partial v^1) X^2
+  (\partial L^{(x_s,v)})(\partial v^2) X^1 = 
a^2 T/\mathcal{N}$. Hence, the insertion of this 
relationship into~(\ref{id4a}) allows us to deduce
\be 
\label{C2a}
C_{2} = \Pi_1 X^2 + \Pi_2 X^1 - \frac{a^2 T}{\mathcal{N}} = 0.
\ee
The constraints~(\ref{C1a}) and~(\ref{C2a}) are nothing but 
the primary constraints arising from an Ostrogradski-Hamilton 
treatment of the original second-order Lagrangian 
function~(\ref{lag0}). In fact, on physical grounds, the 
zero-modes $\lambda^A_{(1)}$ and $\lambda^A_{(2)}$ correspond 
to the velocity and the normal vectors, respectively, associated 
to the brane-like universe under study~\cite{ostro2009}. Then, 
rather than using~(\ref{H1}) and~(\ref{H2}), we will instead 
use $C_1$ and $C_2$ as the new Hamiltonian constraints which 
contain the same information of these constraints but, 
as we will see below, it results adequate to reproduce the 
correct analysis for the gauge symmetries of the theory.
Also, these constraints are fundamental within the naive
quantization program in order to recover the correct 
Wheeler-DeWitt equations in our cosmological setup,
as described in detail in~\cite{ostro2009}.

\sk
Continuing with the iterative procedure for generating further 
constraints, we turn to impose $dC_{1,2}= \{ C_{1,2}, 
H'_{\alpha'} \}^\bullet \,dt^{\alpha'} = 0$. According to the 
functional dependence of~(\ref{C1a}) and~(\ref{C2a}) we have that
\beq
\{ C_{1,2} , H'_{\alpha'} \}^\bullet  &=& (M^{-1})^{A'B'}
\frac{\partial C_{1,2}}{\partial z^{A'}}
\frac{\partial H'_{\alpha'}}{\partial z^{B'}}.
\label{id20}
\eeq
This helps to find  
\beq 
d C_1 &= - \left( \pi_1 \,X^1 + \pi_2\,X^2 + 
\mathcal{N}a^3 \bar{\Lambda}^2 - \frac{aT^2}{\mathcal{N}} 
\right)\,dt^0,
\n
\\
d C_2 &= - \left( \pi_1 \,X^2 + \pi_2\,X^1 \right)\,dt^0.
\n
\eeq
In arriving to these expressions we have used that $\{ C_{1,2}, 
{H}'_{{\alpha}} \}^\bullet = 0$. As a result, we have two new Hamiltonian
constraint functions
\beq
C_3 & = \pi_1 \,X^1 + \pi_2\,X^2 + 
\mathcal{N}a^3 \bar{\Lambda}^2 - \frac{aT^2}{\mathcal{N}} 
= 0,
\label{H3}
\\
C_4 & = \pi_1 \,X^2 + \pi_2\,X^1 = 0.
\label{H4}
\eeq
These correspond to the secondary constraints in the 
Ostrogradski-Hamilton framework. Likewise by imposing 
$d C_{3,4} = 0$ and guided by~(\ref{id20}), 
when considering the functional dependence of~(\ref{H3}) 
and~(\ref{H4}), we deduce that
\beq
dC_{3}&= 0,
\label{condition7}
\\
d C_{4} &= 
\left( \frac{aT^2 \Phi}{\mathcal{N}^3} \right) \mathcal{E} 
\,dt^0
= 0,
\label{condition8}
\eeq
where $\mathcal{E}$ is nothing but the equation of 
motion~(\ref{eom1}) which should not be 
considered as a new constraint.

\sk
The Hamiltonians $C_i$, with $i=1,\ldots,4$, determine a 
non-involutive set of constraints. Indeed, the GPB  
among the $C_i$ and $H'_\alpha$ reads
\be
\begin{array}{ll}
\{ C_1, C_2 \}^\bullet  = 0 &
\qquad \qquad \quad
\{ C_2 , C_3 \}^\bullet = - C_4,
\\
\{ C_1, C_3 \}^\bullet  = - C_3 &   
\quad \qquad \qquad 
\{ C_2 , C_4 \}^\bullet = - C_3 - \bar{\Phi},
\\
\{ C_1, C_4 \}^\bullet  = - C_4  & 
\quad \qquad \qquad 
\{ C_3 , C_4 \}^\bullet =  - \bar{\Theta},
\\
\hspace{-0.05cm}\{ C_i, H'_\alpha \}^\bullet = 0 & \qquad 
\qquad \quad i = 1,2,3,4.
\end{array}
\label{algebra1}
\ee
where $\bar{\Theta}:= (T/\mathcal{N})\,\Theta$ and 
$\bar{\Phi}:= (a/\mathcal{N})\,\Phi$ with $\Theta$ and 
$\Phi$ given by~(\ref{Theta}) and~(\ref{Phi}). 
Therefore, we note that under the GPB we do not recover 
a closed Lie algebra. To ensure integrability of the 
system it is mandatory to redefine these constraints by 
constructing suitable combinations. The Hamiltonians defined by
\beq
\mathsf{H}'_3 &:=& \frac{1}{2}C_1,
\label{f1}
\\
\mathsf{H}'_4 &:=& \bar{\Theta} \,C_2 - \bar{\Phi}\, C_3,
\label{f2}
\eeq
reconstruct the GPB's~(\ref{algebra1}) as follows
\beq 
\{ \mathsf{H}'_3, \mathsf{H}'_4 \}^\bullet &=& -\mathsf{H}'_4,
\label{vira0}
\\
\{ C_2, C_4 \}^\bullet &=& - C_3 -  \bar{\Phi} = -  \bar{\Phi},
\eeq
where the last identity only holds in the constraint surface. 
Accordingly, $\mathsf{H}'_3$ and $\mathsf{H}'_4$, form an 
involutive set of Hamiltonians so that they should be 
considered as generators of the dynamics of the system in 
addition to $H'_\alpha$. This entails their incorporation, 
with some related parameters $t^{\underline{\alpha}}$, to the 
fundamental differential to be constructed. On the other 
hand, $C_2$ and $C_4$ form a non-involutive set of 
constraints and should be treated on an equal footing to 
$H'_{A'}$. This fact leads to introduce the matrix 
$m_{\bar{A}\bar{B}} := \{ C_{\bar{A}}, C_{\bar{B}}\}^\bullet$ 
with $\bar{A},\bar{B} =2,4$, necessary to redefine the 
GPB. Explicitly, this matrix and its inverse are
\be
(m_{\bar{A}\bar{B}}) =  \bar{\Phi}
\left(
\matrix{
0 & -1
\cr
1 & 0
}
\right)
\qquad \qquad
(m^{-1})^{\bar{A}\bar{B}} = \frac{1}{ \bar{\Phi}}
\left(
\matrix{
0 & 1
\cr
-1 & 0
}
\right),
\ee
respectively. Therefore, it turns out that the right 
evolution in the reduced phase space is provided by the 
fundamental differential
\beq  
dF &= \{ F,  H'_{\alpha'}\}^*\, dt^{\alpha'} + \{ F, 
\mathsf{H}_{\underline{\alpha}}\}^*\, dt^{\underline{\alpha}}
\quad\qquad \alpha' = 0,1,2;\,\,\,\underline{\alpha} = 3,4;
\label{dF6}
\eeq
where the final GPB is given by
\be 
\{ F, G\}^* = \{ F, G\}^\bullet - \{F, C_{\bar{A}} \}^\bullet (m^{-1})^{\bar{A}
\bar{B}} \{ C_{\bar{B}}, G \}^\bullet .
\label{gpb}
\ee
This GPB possesses the usual properties similar to those
of the PB. It is crucial to notice that for the HJ approach 
developed for this geodetic cosmological brane model, the 
former $\{ F,G\}^\bullet$, provided by~(\ref{DB2}), gets modified 
so GPB~(\ref{gpb}) should be used instead. Note that this 
final GPB was introduced only to remark the role played by 
the appropriate involutive Hamiltonians of the theory, namely, 
$\mathsf{H}'_3$ and $\mathsf{H}'_4$ which, as mentioned before, 
are the correct generators of the dynamics of the system. 
By a mere relabeling of the constraints $\mathsf{H}'_3 
\rightarrow L_0$ and $\mathsf{H}'_4 \rightarrow L_1$, it 
becomes clear from~(\ref{vira0}) that
\be 
\{ L_n, L_m \}^* = (n-m)L_{n+m} , \qquad \qquad n,m = 0,1.
\ee
This represents a Virasoro truncated 
algebra~\cite{biswajit2013,ho2003,Qmodified2014}. It should be 
remarked that this algebra is valid just on the constraint 
surface. We thus conclude that reparameterization invariance 
symmetry of the RT cosmological brane-like model is equivalent 
to a 2-dimensional conformal gravity described by this peculiar 
type of algebra.

\sk
Once we have at our disposal~(\ref{gpb}), the non-vanishing
fundamental generalized brackets of the theory are
\be 
\fl
\begin{array}{lll}
\{ t, T \}^* = - \frac{\mathcal{N}A^2}{a\Phi} & \quad \,\,\,\{a , T \}^* 
= - \frac{\mathcal{N}TA}{a \Phi} & \quad \{ T, \Pi_A \}^* = - 
\frac{\mathcal{N} \pi_t A}{a {\Phi}}
\\
%\hspace{-0.3cm}
\{ t, A \}^* = -\frac{\mathcal{N} T A}{a \Phi}
& \quad \,\,\,\{a , A \}^* 
= - \frac{\mathcal{N}T^2}{a \Phi}
& \quad \{ A, \Pi_T \}^* = - \frac{\mathcal{N} \pi_a T }{a \Phi}
\\
\{ t, \pi_t \}^* = 1 &  \quad \,\,\,\{a , \pi_a \}^* 
= 1 - \frac{2T^2}{\Phi} & \quad  \{ A, \Pi_A \}^* = 1 - 
\frac{\mathcal{N} \pi_t T }{a \Phi}
\\
\{ t, \pi_a \}^* = - \frac{2TA}{\Phi} &  \quad \,\,\,\{a , \Pi_T \}^* 
= \frac{\mathcal{N}(\Pi_A T - \Pi_T A)}{a\Phi} &
\quad \{ \pi_a, \Pi_T \}^* = - \frac{2\pi_a T}{\Phi}
\\
\{ t, \Pi_T \}^* =  \frac{\mathcal{N} \Pi_A A}{a\Phi} 
+ \frac{a A^3}{\mathcal{N}^2 \Phi} & \quad \,\,\,\{a , \Pi_A \}^* 
= \frac{2\mathcal{N}\Pi_T T}{a\Phi}  & 
\quad \{ \pi_a, \Pi_A \}^* = - \frac{2\pi_t T}{\Phi}
\\
\{ t, \Pi_A \}^* =  \frac{2\mathcal{N} \Pi_T A}{a\Phi} 
 & \quad\,\, \{ T, \Pi_T \}^* = 1 - 
\frac{\mathcal{N} \pi_a A}{a \Phi} & \quad
\{ \Pi_T, \Pi_A \}^* = \frac{\pi_t\,a}{\Phi} 
\end{array}
\ee
Therefore, we find that $(t,\pi_t)$ is the unique canonical 
pair of the theory under the final GPB structure. This fact 
indicates the presence of only one physical degree of freedom.

%%%%%%%%%%%%%%%%%%%%%%%%%%%%%%%%%%%%%%%%%%%%%%%%%%%
\subsection{Characteristic equations}
%%%%%%%%%%%%%%%%%%%%%%%%%%%%%%%%%%%%%%%%%%%%%%%%%

In order to extract physical information we focus first 
on the characteristic equations arising from~(\ref{charact3}).
The first set of equations, taking into account~(\ref{ham0})
and~(\ref{matrix2}), is
\be 
\begin{array}{ll}
dz^1 = dt = X^1\,d\tau & \qquad \qquad
dz^3 = dT = v^1\,d\tau,
\\
dz^2 = da = X^2\,d\tau,
& \qquad \qquad
dz^4 = dA = v^2\,d\tau.
\end{array}
\label{dz1a}
\ee
These reproduce the definitions given in~(\ref{id2}). 
Similarly, the second set of characteristic equations 
turns out to be
\numparts
\beq
dz^5 = d\pi_1 &= 0 
\quad \quad\qquad \qquad \qquad \,\,\Longrightarrow \qquad
\pi_1 = \mbox{const.} =: - \Omega,
\label{dz5}
\\
dz^6 = d\pi_2 &= \frac{\partial L^{(x_s,v)}}{\partial 
x^2}\,dt^0 
\quad \qquad \quad\,\,\, \Longrightarrow \qquad
\frac{\partial L^{(x_s,v)}}{\partial x^2} - \frac{d\pi_2}{d\tau}
= 0,
\label{dz6}
\\
dz^7 = d\Pi_1 &= \left( \frac{\partial L^{(x_s,v)}}{\partial X^1}
- \pi_1 \right)dt^0
\quad \Longrightarrow \qquad
\pi_1 =  \frac{\partial L^{(x_s,v)}}{\partial X^1} - 
\frac{d \Pi_1}{d\tau},
\label{dz7}
\\
dz^8 = d\Pi_2 &= \left( \frac{\partial L^{(x_s,v)}}{\partial X^2}
- \pi_2 \right)d t^0 
\quad \Longrightarrow \qquad
\pi_2 =  \frac{\partial L^{(x_s,v)}}{\partial X^2} - 
\frac{d \Pi_2}{d\tau}.
\label{dz8}
\eeq
\endnumparts
In order to discuss these results we first note 
that, from the HJ viewpoint, expressions~(\ref{dz7})
and~(\ref{dz8}) fix the value of the Lagrange 
multipliers $\pi_1$ and $\pi_2$ through the equations 
of motion. In passing, the remaining Lagrange multipliers
$\Pi_1$ and $\Pi_2$ have been fixed by~(\ref{H1}) and~(\ref{H2}).
Explicitly, they are given by
\beq 
\Pi_1 & = - \frac{a^2 T A}{\mathcal{N}^3} 
\qquad\qquad\qquad \pi_1  = \frac{aT}{\mathcal{N}^3} 
\left[ A^2 + \mathcal{N}^2 (1 - a^2 \bar{\Lambda}^2) \right],
\label{Pi1}
\\
\Pi_2 & =  \frac{a^2 T^2}{\mathcal{N}^3} 
\qquad\qquad\qquad\quad\, \pi_2 = - \frac{aA}{\mathcal{N}^3} 
\left[ A^2 + \mathcal{N}^2 (1 - a^2 \bar{\Lambda}^2) \right].
\label{Pi2}
\eeq
In second place, from~(\ref{dz5}), we note that $\pi_1$ 
becomes a constant as a consequence of the invariance under reparametrizations
of the RT cosmological model. Finally,~(\ref{dz6}) is nothing
but the equation of motion governing the evolution of this brane-like
universe as it may be easily transformed into the Euler-Lagrange form 
\be
\frac{d^2\,}{d\tau^2}\frac{\partial L^{(x_s,v)}}{\partial 
v^2}-\frac{d\,}{d\tau}\frac{\partial L^{(x_s,v)}}{\partial 
X^2} + \frac{\partial L^{(x_s,v)}}{\partial x^2} = 0,
\ee
by direct substitution of relations~(\ref{H2}) and~(\ref{dz8}).
It should be remarked that this expression is totally equivalent
to the eom provided by~(\ref{eom1}). Incidentally, we also note  
from~(\ref{Pi1}) and~(\ref{Pi2}) that $\pi_2 = -(A/T)\pi_1$.

\sk
On the other hand, when considering evolution along the complete
set of parameters, by using~(\ref{dF6}), we are able to find that 
the characteristic equations become
\be 
\begin{array}{ll}
dt = T d \tau - \bar{\Phi}\,T\, dt^4 & \qquad \quad
dT = v^1\,d\tau + \frac{1}{2}T\,dt^3 + \bar{\Theta}A\,dt^4
\\
da = A d \tau - \bar{\Phi}\,A \,dt^4
& \qquad \quad
dA = v^2\,d\tau + \frac{1}{2}A\,dt^3 + \bar{\Theta}\,T\,dt^4
\end{array}
\ee
and
\be 
\fl
\begin{array}{ll}
d\pi_t = 0 & 
\qquad
d\Pi_T = \left( \frac{\partial L^{(x_s,v)}}{\partial T} - \pi_t 
\right) d\tau - \frac{1}{2}\Pi_T \,dt^3 - \frac{\partial 
\mathsf{H}'_4}{\partial T} dt^4,
\\
d\pi_a = \frac{\partial L^{(x_s,v)}}{\partial a} d\tau
- \frac{\partial \mathsf{H}'_4}{
\partial a} dt^4
& \qquad 
d\Pi_A = \left( \frac{\partial L^{(x_s,v)}}{\partial A} - \pi_a
\right) d\tau - \frac{1}{2}\Pi_A \,dt^3 - \frac{\partial 
\mathsf{H}'_4}{\partial A} dt^4.
\end{array}
\ee
From these, we readily observe that the time evolution
of the coordinates is in agreement with that expressed
by (\ref{dz1a}-\ref{dz8}). The additional contributions come 
from the evolution along the parameters $t^3$ and $t^4$, 
which are related to the transformations of the system 
at a fixed time and that remain in the reduced phase space; 
that is, they are associated to the gauge transformations 
of the theory that will be described in short.

%%%%%%%%%%%%%%%%%%%%%%%%%%%%%%%%%%%%%%%%%%%%%%%%%%%%%%%%%%%%%%%
\subsection{The gauge transformations}
\label{sub:3d}
%%%%%%%%%%%%%%%%%%%%%%%%%%%%%%%%%%%%%%%%%%%%%%%%%%%%%%%%%%%%%%

Having at our disposal the Hamiltonians, $H'_{\alpha'}$ and 
$\mathsf{H}'_{\underline{\alpha}}$ generating
the dynamics in the RT cosmology along the 
directions of the parameters $(t^\alpha, t^{\underline{\alpha}})$, 
we are able to construct the gauge generator function as
dictated by~(\ref{G})
\be 
G := H'_\alpha \,\delta t^{\alpha} + \mathsf{H}'_{\underline{\alpha}}
\,\delta t^{\underline{\alpha}}
\qquad \qquad \alpha =1,2, \,\,\,\,\, \underline{\alpha} = 3,4.
\ee
In this sense, when considering~(\ref{gpb}), the infinitesimal 
gauge transformations are
\be 
\delta_G z^A = \{ z^A, G \}^* = \{ z^A, H'_\alpha \}^* \,
\delta t^\alpha + \{ z^A , \mathsf{H}'_{\underline{\alpha}}
\}^*\,\delta t^{\underline{\alpha}}.
\label{deltaGzA}
\ee
These transformations leave invariant the action functional~(\ref{action})
with the Lagrangian given by~(\ref{L0}). Taking into account 
the functional dependence of $H'_\alpha$ we have that $\{ z^A, 
H'_\alpha \}^* = \delta^A{}_\alpha$, thus for the model under 
study the gauge transformations~(\ref{deltaGzA}) become
\beq 
\delta_G \,t & 
= \frac{\partial 
\mathsf{H}'_{\underline{\alpha}}}{\partial \pi_t}\,
\delta t^{\underline{\alpha}} = - \bar{\Phi} \,T\,\delta t^4,
\label{deltat}
\\
\delta_G \,a & 
= \frac{\partial 
\mathsf{H}'_{\underline{\alpha}}}{\partial \pi_a}\,
\delta t^{\underline{\alpha}} = - \bar{\Phi} \,A\,\delta t^4,
\label{deltaa}
\\
\delta_G T & 
= \frac{\partial \mathsf{H}'_{\underline{\alpha}}}{\partial \Pi_T}\,
\delta t^{\underline{\alpha}} = \frac{1}{2} T \,\delta t^3 
+ \bar{\Theta}\,A\, \delta t^4,
\label{deltaT}
\\
\delta_G A & 
= \frac{ \partial \mathsf{H}'_{\underline{\alpha}}}{\partial \Pi_A}\,
\delta t^{\underline{\alpha}} = \frac{1}{2}A\,\delta t^3 
+  \bar{\Theta}\,T\,\delta t^4.
\label{deltaA}
\eeq

\sk
It follows that the variation of the Lagrangian~(\ref{eqL}), 
taking into account both~(\ref{ham0}),~(\ref{KA}) and~(\ref{matrix2}), 
can be written in the form $\delta L = 
-( \dot{\pi}_t + \partial H^{(x_s,v,\pi^s)}/\partial t) \delta t
- (\dot{\pi}_a + \partial H^{(x_s,v,\pi^s)}/\partial a)\delta a
- (\dot{\Pi}_T + \partial H^{(x_s,v,\pi^s)}/\partial T)\delta T
- (\dot{\Pi}_A + \partial H^{(x_s,v,\pi^s)}/\partial A)\delta A
+ (d/d\tau) (\pi_t \delta t + \pi_a \delta a + \Pi_T \delta T
+ \Pi_A \delta A)$. Bearing in mind the  independence on the 
parameter $t$ of the rest of the terms occurring in the model, as 
well as the fact that $\pi_t$ is a constant, the variation of the 
Lagrangian becomes 
\beq 
\delta L & = \left(\frac{\partial L^{(x_s,v)}}{\partial a}
- \dot{\pi}_a \right) \delta a
+  \left(\frac{\partial L^{(x_s,v)}}{\partial T}
- \dot{\Pi}_T - \pi_t \right) \delta T
\n
\\
&+ \left(\frac{\partial L^{(x_s,v)}}{\partial A}
- \dot{\Pi}_A - \pi_a \right) \delta A
+ \frac{d}{d\tau} (\pi_t \delta t + \pi_a \delta a + \Pi_T \delta T
+ \Pi_A \delta A),
\label{deltaLE}
\eeq 
where~(\ref{ham0}) has been considered. 
On the other hand, by considering the definitions $\epsilon_2 (\tau) 
:= \bar{\Phi} \,\delta t^4 $ and $2\epsilon_1 := \delta t^3$, the 
transformations~(\ref{deltat}-\ref{deltaA}) reduces to
\beq 
\delta_G \,t &=  - T\,\epsilon_2 ,
\label{deltat2}
\\
\delta_G \,a &=  - A\,\epsilon_2,
\label{deltaa2}
\\
\delta_G T &=  \epsilon_1 \,T + \left(\frac{T\,
{\Theta}}{a{\Phi}}\right) \epsilon_2\,A,
\label{deltaT2}
\\
\delta_G A & = \epsilon_1 \,A + \left(\frac{T\,
{\Theta}}{a{\Phi}}\right) \epsilon_2\,T.
\label{deltaA2}
\eeq
By plugging these transformations in the variation~(\ref{deltaLE}), 
after a lengthy but straightforward computation, one finds that
\beq
\delta L & = - C_3\,\epsilon_1 - \left( \frac{T \Theta}{a \Phi}
\right)C_4 \,\epsilon_2 - \left( \frac{a AT^3 \Phi}{\mathcal{N}^5}
\right) \mathcal{E}\,\epsilon_2
\n
\\
&+ \frac{d}{d\tau} \left\lbrace  2\mathsf{H}'_3 \,\epsilon_1
+\frac{N}{a \Phi} \mathsf{H}'_4 \,\epsilon_2 
+ \frac{a}{\mathcal{N} \Phi} \left[ T^2 \Theta 
- (T^2 - \mathcal{N}^2 a^2 \bar{\Lambda}^2 ) \Phi \right] 
\epsilon_2 \right\rbrace,
\label{deltaLE3}
\eeq
where we recognise  the eom $\mathcal{E}$ provided by~(\ref{eom1}). 
Therefore, under the variations~(\ref{deltat2}-\ref{deltaA2}) the change
induced in the Lagrangian~(\ref{L0}) left this invariant 
whenever $\epsilon_2 (\tau)$ vanishing at the extrema located 
at $\tau = \tau_1$ and $\tau = \tau_2$.
To be more specific, the action~(\ref{action}) with the 
Lagrangian~(\ref{lag0}) is left invariant under the gauge transformations~(\ref{deltat2}-\ref{deltaA2}) where one must 
have in mind the following parameter relationship
\be 
\epsilon_1 = - \lambda (\tau)\,\epsilon_2 - \dot{\epsilon}_2,
\ee
where $\epsilon_2$ is subject to the conditions $\epsilon_2 (\tau_1)
= \epsilon_2 (\tau_2) = 0$. It follows then that $\epsilon_2$ 
may be chosen as the independent gauge parameter confirming 
the existence of only one degree of freedom where $\lambda 
(\tau)$ is an arbitrary function. Certainly, this expression can 
be obtained by imposing the standard commutativity requirements 
provided by the usual variational principles, namely,
\be 
\label{eqV}
\frac{d}{d\tau} \delta t = \delta T \qquad \mbox{and} \qquad
\frac{d}{d\tau} \delta a = \delta A \,. 
\ee
We therefore have systematically obtained the gauge symmetries 
from a purely geometrical viewpoint. 

\sk
A couple of comments are in order. The gauge variations $\delta_G \,t$
and $\delta_G\,a$ reflect the presence of the invariance under 
reparametrizations of the model as one can notice from
(\ref{deltat2}) and~(\ref{deltaa2}),
\be
\delta_G \,t = - \epsilon_2 \,T \qquad \mbox{and} \qquad
\delta_G \,a = - \epsilon_2 \,A.
\ee
These preserve the action~(\ref{action}) with $L$ provided
by~(\ref{lag0}) under the transformation $\tau \rightarrow
\tau + \epsilon_2 (\tau)$. Moreover, the gauge variations 
$\delta_G T$ and $\delta_G A$ reflect the presence of an 
inverse Lorentz-like transformation in the velocity sector of 
the coordinates $z^A$, provided by $T$ and $A$, as we can 
observe from~(\ref{deltaT2}) and~(\ref{deltaA2}) 
\be
\delta_G T = \epsilon_1 \left( T + \widetilde{\epsilon}_2
A \right) \qquad \mbox{and}\qquad 
\delta_G A = \epsilon_1 \left( A + \widetilde{\epsilon}_2
T \right),
\ee
with $\widetilde{\epsilon}_2:= (T\Theta/a\Phi)
(\epsilon_2/\epsilon_1)$.

%%%%%%%%%%%%%%%%%%%%%%%%%%%%%%%%%%%%%%%%%%%%%%%%%%%%%%%
\section{On the Hamilton principal function}
%%%%%%%%%%%%%%%%%%%%%%%%%%%%%%%%%%%%%%%%%%%%%%%%%%%%%%%

By virtue of the integrability analysis developed 
for the RT cosmological model, the conditions to obtain 
the Hamilton principal function, $S = S[t^{\alpha'},z^{A'} 
(t^{\alpha'})] = S[\tau, z^{A'}(\tau)]$, are fulfilled. 
First of all, we focus  on the second term on the RHS occurring 
in~(\ref{dS3})
\be 
H_{A'} (M^{-1})^{A'B'} \frac{\partial H^{(x_s,v,\pi^s)}}{\partial 
z^{B'}} = - \left( \pi_1 X^1 + \pi_2 X^2 + \Pi_1 v^1
+ \Pi_2 v^2 \right),
\label{identity2}
\ee
where~(\ref{ham0}),~(\ref{KA}) and~(\ref{matrix2}) have 
been considered. When inserting this into~(\ref{dS3}) we obtain
\be
dS =  \left[ \pi_1 X^1 + \pi_2 X^2 + \Pi_1 v^1
+ \Pi_2 v^2 - H^{(x_s,v,\pi^s)} \right] dt^0.
\nonumber
\ee
By integrating and using once again the relationship~(\ref{ham0}), 
we get
\be 
S = \int L^{(x_s,v)}\,d\tau,
\label{S1}
\ee
which is nothing but the action~(\ref{action}) with $L^{(x_s,v)}$
provided by~(\ref{L0}). We therefore realize that the action 
$\mathcal{S}$ is a solution of the HJPDE, as expected. 
On the other hand, from~(\ref{dS0}) we have
\be 
\label{dS4}
dS =  K_{A'} \,dt^{A'} - H^{(x_s, v, \pi^s)} \, dt^0,
\ee
where~(\ref{def1}) has been considered. 
Now, within the minisuperspace geodetic brane scenario, by 
inserting the $K_A$ terms provided by~(\ref{KA}), and integrating 
we obtain
\be 
S = \int \left[ \pi_1 \,dx^1 + \pi_2\,dx^2 +
\Pi_1 \, dX^1 + \Pi_2 \, dX^2 - H^{(x_s, v, \pi^s)} \, dt^0 \right].
\ee
Bearing in mind that the variables $z^{A'}$ only depend on
$t^0 = \tau$, we find that
\beq 
S &= \pi_1 \,x^1 + \pi_2\,x^2 + \Pi_1 X^1 + \Pi_2 X^2
- \int H^{(x_s, v, \pi^s)} \, dt^0,
\n
\\
&= \pi_t \,t + \pi_a\, a,
\label{S2}
\eeq
where the constraints~(\ref{C1a}) and the fact that $H^{(x_s, v, 
\pi^s)}= 0$  have been used. Indeed, the previous identity coincides 
exactly with the constraint $C_3 = 0$, when one rewrite the 
Lagrangian function $L^{(x_s,v)}$, (\ref{L0}), in terms of the 
quantities $\Pi_T$ and $\Pi_A$, provided by~(\ref{Pi1}) 
and~(\ref{Pi2}). One must bear in mind that this fact implies 
that the evolution is carried out in the reduced phase space. 
In order to extract physical information regarding the semi-classical
quantization of the model, we prefer to write the function $S$
in terms of the constant $\Omega$,~(\ref{dz5}). As noticed
above, $\pi_t = - \Omega$ and $\pi_a = \frac{A}{T}\Omega$, so that
\be 
S(z^{A'}, \tau) = - \Omega \,t + \frac{A}{T}\,\Omega\,a .
\label{S3}
\ee

As it is well known, the HJ equation~(\ref{eq7}) may be 
viewed as the classical limit of quantum field 
equations~\cite{messiah1999}. In particular, the HJ equation 
turns out to be the classical limit of the Schr\"{o}dinger 
equation of non-relativistic quantum mechanics, and hence the 
interest in computing the Hamilton principal function. In such 
approach, the complete wave function is computed by considering 
the ansatz $\Psi (z^{A'}, \tau) = e^{\frac{i}{\hbar} S(z^{A'}, 
\tau)}\psi (z^{A'}, \tau)$. Within the minisuperspace cosmological 
brane scenario, the important quantities are the external time 
$t$ and the scale factor $a$, as one can observe from~(\ref{lag0}) 
whereas the velocities $T$ and $A$ are considered as functions of 
$\tau$. Additionally, in the same context,
the fact that $\pi_t = - \Omega$ arises from the feature that 
the theory is independent of $t$ at the classical level. 
In that sense, for the case under study, from~(\ref{S3}) we note
that the wave function has the ordinary time dependence 
$e^{-i \frac{\Omega}{\hbar}\,t}$, as expected.
Hence, in a semi-classical
approximation,  we have that the brane-like wave functions
acquire the form
\be 
\Psi (t,a;\tau) = \psi (a;\tau) e^{i \left[ \frac{(A/T) 
\Omega}{\hbar} a - \frac{1}{\hbar} \Omega\,t\right]} 
=: \psi (a;\tau) e^{i \left( k_a\,a  - \omega\,t\right)}
\label{psi2}
\ee
where $k_a : = (A\Omega/T)/\hbar$ and $\omega := \Omega/\hbar$.
This represents an outgoing wave whenever we assume that $\Omega >0$.
This result has been obtained from different perspectives mainly based
in the naive quantization supported by the Dirac-Bergmann
theory for classical constrained systems~\cite{ostro2009,davidson1999}.
However, we must emphasize here that within the HJ formalism this result
naturally emerged as a consequence of the integrability conditions 
that allowed us to obtain, in a straightforward manner, the Hamilton 
principal function for geodetic brane cosmology.

%%%%%%%%%%%%%%%%%%%%%%%%%%%%%%%%%%%%%%%%%%%%%%%%%%%%%%%%%%%%%%%%%
\section{Concluding remarks}
\label{sec:6}
%%%%%%%%%%%%%%%%%%%%%%%%%%%%%%%%%%%%%%%%%%%%%%%%%%%%%%%%%%%%%%%%

In this paper we have derived a Hamilton-Jacobi framework
for geodetic brane cosmology. Being a singular theory 
with a linear dependence in the accelerations of the brane,
by judiciously enlarging the configuration space we were able
to properly treat this type of cosmology within the HJ 
framework. In consequence, we obtained a set of HJPDE instead of a single 
HJ equation, as it occurs for the case of regular theories. When identifying 
the complete set of involutive Hamiltonians that play the 
role of the distinct generators of the evolution of the system, and 
following the integrability conditions, we realize that it was 
compulsory to properly modify the original GPB,~(\ref{DB2}) by a slightly 
modified one,~(\ref{gpb}), in order to obtain the right 
dynamical evolution of the system in the reduced phase space. 
As discussed above, this modified GPB was consequently introduced 
in order to consider the appropriate involutive Hamiltonians of 
the theory that were related to the correct generators of the 
dynamics.  As expected, the non-involutive Hamilton constraint 
functions were eliminated as dynamical generators by this modified 
GPB. The adequate set of involutive constraints, $(H'_\alpha, \mathsf{H}'_{\overline{\alpha}})$, 
allowed us to become aware of the integrability of the HJ equations.
We also note that there is a subset of the Hamiltonian generators 
that under the modified GPB structure closes as
a truncated Virasoro algebra which, as discussed in the 
literature~\cite{biswajit2013,ho2003,Qmodified2014}, may codify the 
information of a 2-dimensional conformal symmetry that results 
inherent to the model of our interest here. This last issue is still 
under active study.  Further, the HJ scheme provided a purely 
geometrical approach that perfectly adjusts to construct the
gauge symmetries of the theory. Indeed, we were able to 
straightforwardly built the generator of gauge transformations 
in terms of the generators of the dynamical evolution along each 
of the directions of the independent parameters of the theory. 
Finally, by obtaining the generic form of the Hamiltonian principal 
function for the RT cosmology, we have outlined in a simpler manner 
the generalities of the wave function within the context of the 
semi-classical approximation. This last result may be compared with 
the equivalent results commonly found in the literature where the 
wave function is encountered by invoking barely geometrical principles, 
as those involved within our present prescription. We believe 
that the extensions of the  HJ scheme developed here are quite 
natural to be implemented within the generic context of 
the so-called affine in acceleration theories~\cite{affine2016}. 
Work in this direction will be reported elsewhere.

%%%%%%%%%%%%%%%%%%%%%%%%%%%%%%%%%%%%%%%%%%%%%%%%%%%%%%%%%%%%%%%%%%%%%%%%%
\section*{Acknowledgments}
%%%%%%%%%%%%%%%%%%%%%%%%%%%%%%%%%%%%%%%%%%%%%%%%%%%%%%%%%%%%%%%%%%%%%
The authors thank R. Cordero for stimulating discussions and 
valuable comments. ER thanks ProdeP-M\'exico, CA-UV-320: Algebra, 
Geometr\'\i a y Gravitaci\'on. Also, ER thanks the Departamento de
F\'\i sica de la Escuela Superior de F\'\i sica y Matem\'aticas
del Instituto Polit\'ecnico Nacional, M\'exico, where 
part of this work was developed during a sabbatical leave. AAS 
thanks support from a CONACyT-M\'exico doctoral fellowship.  
AM would like to acknowledge financial support from CONACYT-Mexico under
project CB-2014-243433. This work was partially supported by SNI (M\'exico).

%%%%%%%%%%%%%%%%%%%%%%%%%%%%%%%%%%%%%%%%%%%%%%%%%%%%%%%%%%%%%%%%%%%
%%%%%%%%%%%%%%%%%%%%%%%%%%%%%%%%%%%%%%%%%%%%%%%%%%%%%%%%%%%%%%%%%

\end{document}